\documentclass[journal]{IEEEtran}
\usepackage[numbers,sort&compress]{natbib}
\usepackage{array}
\usepackage{amsfonts,amsmath,amsthm}
\usepackage{amssymb}
\usepackage{subfigure,graphicx}
\usepackage{algorithm}
\usepackage{algorithmicx}
\usepackage{algpseudocode}
\usepackage{amsfonts}
\usepackage{epstopdf}
\usepackage{color}
\usepackage{caption}

\begin{document}

\title{OL4EL: Online Learning for Edge-cloud Collaborative Learning on Heterogeneous Edges with Resource Constraints}

\author{Qing Han, Shusen Yang, Xuebin Ren, Cong Zhao, Jingqi Zhang, Xinyu Yang

\thanks{Qing Han, Shusen Yang, Xuebin Ren, Jingqi Zhang, and Xinyu Yang are all with the National Engineering Laboratory for Big Data Analytics (NEL-BDA) at Xi'an Jiaotong University. Cong Zhao is with the Department of Computing at Imperial College London.}

}


\maketitle

\begin{abstract}
  Distributed machine learning (ML) at network edge is a promising paradigm that can preserve both network bandwidth and privacy of data providers.
  However, heterogeneous and limited computation and communication resources on edge servers (or edges) pose great challenges on distributed ML and formulate a new paradigm of Edge Learning (i.e. edge-cloud collaborative machine learning).
  In this article, we propose a novel framework of `learning to learn' for effective Edge Learning (EL) on heterogeneous edges with resource constraints.
  We first model the dynamic determination of collaboration strategy (i.e. the allocation of local iterations at edge servers and global aggregations on the Cloud during collaborative learning process) as an online optimization problem to achieve the tradeoff between the performance of EL and the resource consumption of edge servers.
  Then, we propose an Online Learning for EL (\textsf{OL4EL}) framework based on the budget-limited multi-armed bandit model.
  \textsf{OL4EL} supports both synchronous and asynchronous learning patterns, and can be used for both supervised and unsupervised learning tasks.
  To evaluate the performance of \textsf{OL4EL}, we conducted both real-world testbed experiments and extensive simulations based on docker containers, where both Support Vector Machine and K-means were considered as use cases.
  Experimental results demonstrate that \textsf{OL4EL} significantly outperforms state-of-the-art EL and other collaborative ML approaches in terms of the trade-off between learning performance and resource consumption.
\end{abstract}

\IEEEpeerreviewmaketitle

\section{Introduction} \label{sec:introduction}

With the proliferation of IoTs and 5G technologies, the number of sensors and smart devices served by edge networks has been exploding, where an exponentially increasing amount of data are generated and required to be understood.
Machine Learning (ML), as one of the most promising solutions to the big data utilization, is being applied to a broadened spectrum of fields like augmented reality, autonomous driving, smart manufacturing~\cite{hu2015mobile} \emph{etc.}
However, the transmission of such big data to the Cloud for centralized ML is demonstrated to be prohibitive considering both the burden of the backbone network and the concern of data privacy in practice.
To address this issue, the notion of collaborative ML within the emerging edge computing paradigm~\cite{taleb2017multi} has been proposed, which aims at achieving agile, fast, and cost-effective ML through collaborative training among distributed devices or servers at the network edge (\emph{e.g.} smartphones, autonomous vehicles, IoT gateways, micro data centers~\cite{hu2015mobile}), with the coordination on the Cloud.
Such a distributed collaborative ML approach demonstrates great advantages in terms of bandwidth saving, delay reduction, and privacy preservation, and is attracting increasing interests from both academia~\cite{kairouz2019advances} and industry~\cite{bonawitz2019towards}.
As a representative example, Federated Learning (FL) proposed by Google~\cite{bonawitz2019towards} trains ML models among myriad smartphones without inspecting their data, thus protecting users' privacy.

Different from the cross-device FL~\cite{kairouz2019advances} utilizing massive end devices (\emph{e.g.} smartphones, tablets), we consider a cross-silo FL~\cite{kairouz2019advances} with edge servers (or edges, \emph{e.g.} geo-distributed datacenters, IoT gateways, 5G connected cars~\cite{hu2015mobile}) that provides authenticated access to efficient backbone or core networks. We refer to it as edge-cloud collaborative ML or Edge Learning (EL). Consider two typical scenarios of EL below:
\begin{itemize}
\item \textbf{AI Self-driving Cars.} In Internet of Vehicles, tons of timely driving data from self-driving cars can be collaboratively trained to improve AI self-driving practice~\cite{hu2015mobile}. However, it is estimated that a self-driving car that runs eight hours a day would produce at least 40TB of data. That would cost a huge amount of network traffic and battery energy~\cite{ning2019mobile}. For the sake of safety, cars with different computing speeds update the status information in an asynchronous manner for fast response.

\item \textbf{Edge Cloud based AI.} With the emergent edge computing, numerous micro datacenters will sprout up at the network edge to form the edge clouds~\cite{taleb2017multi}. Edge cloud based AI services depend on collaborative ML for geo-distributed datacenters that vary greatly in computing capacities. As FaaS (Function-as-a-Service) or serveless computing techniques~\cite{baldini2017serverless} being used, pricing of edge cloud services is often based on the immediate resources consumed, such as time allocated to the services.
\end{itemize}

In both EL scenarios, the heterogeneity (\emph{i.e.} varied computation capabilities) and resources (\emph{e.g.} energy capacities of cars or monetary budgets in edge clouds) of edge servers greatly impact the EL performance, and even the service sustainability.
Therefore, it is essential to seek the cost-effective EL under the \textit{heterogeneous edges} with different \textit{resource constraints}.

Most studies on similar problems focus on synchronous parallelism where homogeneous nodes (end devices or edge servers) update models simultaneously.
For instance, considering the limited bandwidth, \cite{mcmahan2016communication} proposed the synchronous FedAvg framework that uses additional computations on distributed nodes to reduce communications in FL.
\cite{konevcny2016federated} and \cite{hsieh2017gaia} explored the compression techniques and threshold-based updates, respectively, to save up communications.
Recently, to tackle the straggler effect caused by increasingly heterogeneous nodes, asynchronous patterns have also received considerable attentions.
For example, \cite{mcmahan2014delay} provided an efficient asynchronous algorithm by adaptively tuning and revising the learning rates.
\cite{lian2017asynchronous} proposed an asynchronous framework for decentralized stochastic gradient decent.
However, all these studies concentrate on alleviating the communication cost while neglecting the computation overhead, thus failing to consider the learning efficiency in terms of overall resource constraints.
The most relevant work to this article is \cite{wang2018edge}, which studied how to optimize the learning with limited resources for both computation and communications by theoretically analyzing the convergence rate of gradient-descent based distributed learning.
However, the theoretical analysis is built on the synchronous setting and can not contribute to asynchronous EL with highly heterogeneous edge servers.

Therefore, there still lacks a systematic discussion on the cost-effective EL approach for heterogeneous edge servers with resource constraints, which faces several challenges:
\begin{itemize}
\item \textbf{Edge heterogeneity}. Most predominating synchronous algorithms are inefficient in confronting heterogeneous edge servers due to the severe straggler effect.
\item \textbf{Model complexity}. For asynchronous solutions, it is difficult to mathematically model the relationship between the EL model accuracy and edge resource consumption.
\item \textbf{System dynamics}. Both training data and heterogeneous edge resource consumption can be time-varying, which will cause unpredictable impacts on EL.
\end{itemize}

Considering above challenges, we model the decision strategy for effective EL as an online optimization problem.
Then, we develop an Online Learning for Edge Learning (\textsf{OL4EL}) solution based on Multi-Armed Bandit (MAB) method to adaptively pursue the optimal tradeoff between the EL utility and the edge resource consumption.
To the best of our knowledge, \textsf{OL4EL} is the first OL-based EL algorithm framework, and the first EL framework for heterogeneous edges under resource constraints.
The contributions are as follows:

\begin{enumerate}
    \item
    We model the dynamic decision of collaborative learning strategy (the frequency sequence of local iterations at the edges and global aggregations at the Cloud) among heterogeneous edges with different resource limitations as an online optimization problem, which is further formulated as a budget-limited MAB problem.
    \item
    We propose an algorithmic framework \textsf{OL4EL} for our specific resource-constrained EL problem, based on the upper confidence bound theories for budget-limited MAB. \textsf{OL4EL} can seek the approximately optimal cost-effective learning strategies on-the-fly.
    Particularly, different algorithms are developed for synchronous and asynchronous EL scenarios with fixed and variable resource consumption rates, respectively.
    \item
    We conducted both real-world testbed experiments and extensive simulations based on docker containers to evaluate the performance of \textsf{OL4EL}.
    Experimental results demonstrate that our approach outperforms existing solutions in terms of the tradeoff between model accuracy and edge resource consumption (e.g., 12$\%$ enhancement on model accuracy under the same resource constraints).
\end{enumerate}

\section{Background and Motivation}

\begin{figure*}
  \centering
  {
  \includegraphics[width=0.9\textwidth]{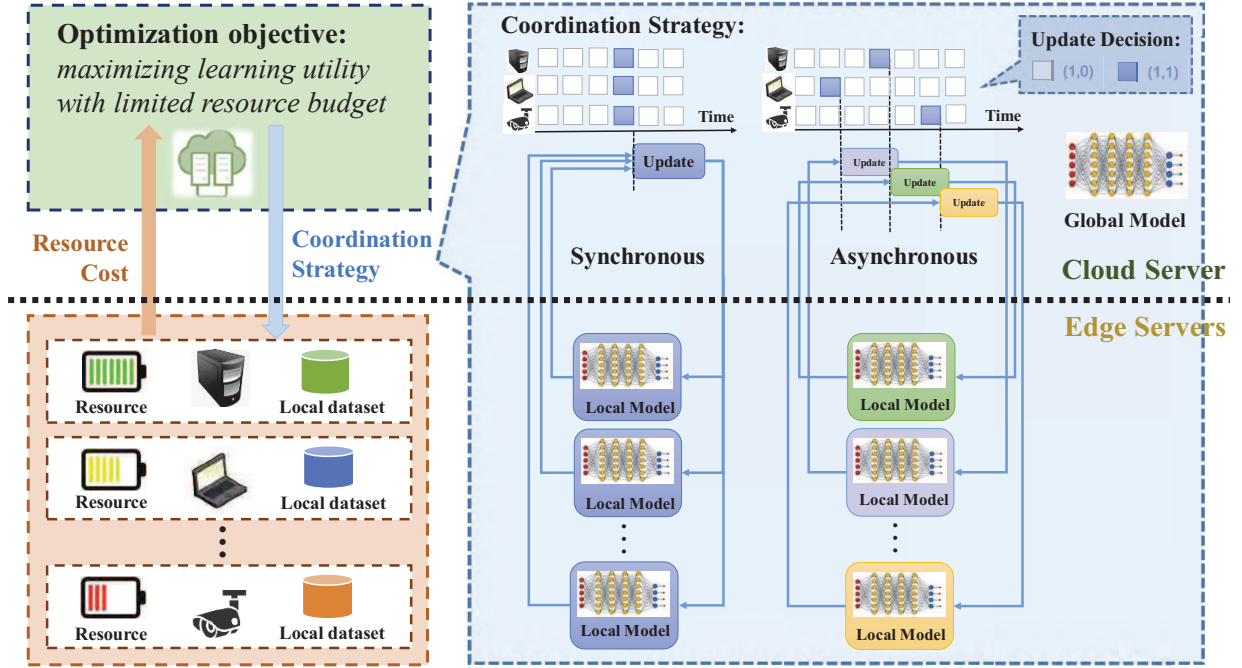}
  }
  \center\caption{Edge-cloud Collaborative Learning Framework. \label{fig: edge-cloud system}}
\end{figure*}

\textbf{Federated Learning} is a typical distributed collaborative ML approach aims to train a high-quality model from massive end devices, which is effective for mitigating the data isolation while preventing data leakage.
Different from the cloud learning that possesses unlimited computing power, end devices are usually subject to limited computation and communication capacities, and considered to be unreliable.
To guarantee sufficient resources for effective training, many practical FL systems require that participating smartphones must meet the strict requirements of being idle, charging, and connected to unmetered network such as WiFi~\cite{bonawitz2019towards}.
Due to unreliability and resource constraints, FL devices often participate in training in an opportunistic way without carrying any state.
Therefore, a crucial problem in FL is to minimize the resource spend of end devices.
However, most proposed FL approaches either merely focus on communication reduction without touching the computation resource, or implicitly favor the synchronous learning architecture for mostly homogeneous devices.

\textbf{Edge Learning} migrates the FL framework from end devices to the edge servers.
Compared with FL that allows massive unreliable devices participate in the learning statelessly, EL focuses on reliable and stateful edge servers that can persistently participate in the entire training until convergence or resource exhaust.
Besides, edge servers in EL are much fewer but more heterogeneous than end devices in FL.
Therefore, simply synchronous learning scheme becomes unfavorable for practical EL settings due to unacceptable latency and unfairness for faster edge servers.
In this article, we aim to investigate cost-effective EL training for stateful but heterogeneous edges with limited computation and communication resources, under both the synchronous and asynchronous learning frameworks.
Considering the dynamics and resource constraints, the problem of cost-effective EL training can be regarded as an online optimization problem that maximizes the model utility among heterogeneous edge servers subject to various resource constraints.

\textbf{Online Learning} is an important domain in machine learning with interesting theoretical properties and practical applications.
In the OL setup, the decision maker repeatedly selects available policies.
Only after the choice can the decision-maker realize the meaning of the chosen strategy and derive practical value.
Even in the case of complete uncertainty about the environment (no hypothesis/inference about any relationship between the strategy and the utility value), there is still an OL algorithm for that state and provides a verifiable guarantee (namely the classical ``no regret'' guarantee).
Recently, OL methods have been used in various network optimization and intelligent decision making.
We are thus inspired to apply ideas and machinery from online learning to optimize the learning strategy of edge learning.

\section{System Model and Problem Definition}\label{Sec:System}

We consider the edge learning process in an edge-cloud collaborative system that consists of one Cloud server and $N$ edge servers with both heterogeneous resource constraints and different local datasets.
Each edge server maintains a local parametric ML model, and the Cloud server maintains a global model.
The goal of the Cloud server is to effectively train the global model by coordinating the \textbf{local iterations} at the edge servers (\emph{i.e.} local model updates based on batches of local data) and the \textbf{global updates} on the Cloud (\emph{i.e.} global model updates based on the local models received from edge servers) under specific resource constraints.
For each global update, we consider two different collaboration manners as shown in Figure \ref{fig: edge-cloud system}:
1) Synchronous manner:
the Cloud server requests all edge servers to upload their local models, and updates the global model by calculating the weighted average of all local models.
2) Asynchronous manner:
the Cloud server only requests one edge server to upload its local model to update the global model.
Then, the latest global model is replied to edge servers who contribute to the global update.

\subsection{Coordination Strategy and Learning Utility}\label{subsec: strategy and utility}

\textbf{Coordination strategy:}
We assume that the learning process is defined on discrete time slots $t=0, 1, 2, ...$, and is initiated by the Cloud server.
When $t=0$, we set the global model randomly.
For each $t>0$, both the local iterations at edges and the global update at Cloud are performed according to the \textit{coordination strategy} maintained by the Cloud server.

For each edge server $i$, we use two binary variables to denote the \textit{decisions} of local iteration and global update respectively.
Specifically, we define $i$'s \textit{update decision} set as $\{(0,0), (1,0), (1,1)\}$, where $(0,0)$ means `neither local iteration nor global update', $(1,0)$ means `local iteration but no global update' and $(1,1)$ means `global update after local iteration'.
Note that the case $(0,1)$ is omitted since that the case `global update without local iteration' should never appear.
Accordingly, we define the \textit{coordination decision} of the Cloud server at slot $t$ is a set consisting of the \textit{update decision} of all edge servers at slot $t$.
Based on definitions above, we define the \textit{coordination strategy} of the Cloud server is the \textit{coordination decision} sequence till slot $t$.

\textbf{Learning utility:}
For each time slot $t$, we measure the performance of the global model as the \textit{learning utility} (the learning utility should be model-specific), which can be treated as a parametric function of the current \textit{coordination decision} and testing set that consists of a negligible amount of raw data uploaded by edge servers.
The utility can be evaluated by the Cloud server only when a global update is conducted, where the testing set is uploaded to the Cloud together with corresponding local models.
We can also measure the \textit{learning utility} according to the difference between the global parameters at current slot $t$ and slot $t-1$.
Specifically, smaller difference means higher utility and vice versa.
For example, in K-means algorithm, we can define the \textit{learning utility} as the negative value of Euclidean distance between the cluster centers learned at two consecutive slots.

\subsection{Resource Costs and Constraints}

In practice, compared with the powerful datacenter-based Cloud server with almost `unlimited' capabilities, distributed edge servers always have different \textit{resource constraints}.
Here, \textit{resource} is a generic notion that refers to the execution overhead (e.g., occupying time, memory, energy or monetary cost) related to computations and communications for the entire edge learning process (including but not limited to extra resource consumption due to possible security protocols for securing EL) at the edge servers.
In particular, we assume that, for each edge server, each local iteration consumes certain amounts of computation resource for local model update, and each global update consumes certain amounts of communication resource for edge-cloud interactions.
Then, during the learning process that combines both local iterations and global updates, the constrained resources of all edge servers may eventually run out, especially when there are large number of data batches generated at the edges.

Often the resource consumptions of both computation and communication are metered under or can be converted into the same measurement (\emph{e.g.} time, energy, monetary cost).
In such a way, the resource cost of each edge server can be simplified as a the sum of the computation (for local iterations) and communication (for global updates) resource costs.
Meanwhile, a total budget for each edge server can be given as its resource constraint.
For the ease of discussion, we stick to such an assumption for the rest of the article.
Additionally, an edge server's specific resource costs for each local iteration and global updates could be either fixed values through the entire learning process or, from the practical perspective, time-varying values considering the system randomness.

\subsection{Problem Definition}

The goal of the Cloud server is to determine the optimal \textit{coordination strategy} that maximizes the average learning utility across the entire learning process that has to be terminated (at slot $T$) before all of resource constraints are consumed.

However, since the training batches from the local dataset of each edge server come with uncertainty at each slot, the relationship between \textit{learning utility} and \textit{resource cost} cannot be explicitly formulated, which makes it difficult for the Cloud server to determine the optimal \textit{coordination strategy}.
Inspired by online optimization theory, during the edge learning process over time-varying training data on edge servers, the Cloud server faces a trade-off between \textit{exploiting current knowledge} to maintain the strategy that has brought the highest learning utility so far and \textit{exploring new strategies} that might bring higher utility in future.

\section{The Online Learning for Edge Learning (OL4EL) Algorithm}\label{sec:Online}

\subsection{Bandit Formulation}

The strategy decision confronting the dilemma between exploration and exploitation is usually formalized as a \textit{bandit problem}~\cite{auer2002finite}, which has various variants.
Considering the edge resource constraints, our problem can be modeled as a \textit{budget-limited bandit problem}~\cite{tran2012knapsack,ding2013multi}.

According to the aforementioned definitions, for each edge server, we define the number of local iterations between two adjacent global updates as its \textit{global update interval}, which is maintained and selected by the Cloud server from an discrete integer set ranging from $1$ to a predefined longest interval.
Its resource cost corresponds the computation resource cost for local iterations plus the communication resource cost for global updates.
For example, suppose an edge server has finished a global update and receives the global update interval of $3$, then it needs to conduct three rounds of local iterations before its next global update and consumes a total resource equals to certain computation resource for three local iterations plus certain communication resource for one global update.
Therefore, the \textit{coordination decision} set at slot $t$ can be transformed as the \textit{global update interval} set.

According to the bandit terminology, for each edge server, we denote each \textit{global update interval} as an \textit{arm}, while the resource consumption and learning utility for each \textit{global update interval} corresponds to the cost and reward of the arm, respectively.
The resource constraint of each edge server corresponds to the budget for the arm cost.
Therefore, based on the above EL models, our specific EL problem of finding the cost-effective coordination strategy to maximize the average learning utility under the resource constraints is mapped into the budget-limited multi-armed bandit problem that seeks the optimal arm sequences to maximizes the average arm reward while keeping the total arm cost no more than given budgets.

\subsection{Multi-armed Bandit Algorithm based Online Learning Mechanism}

\begin{figure}
  \centering
  {
  \includegraphics[width=0.5\textwidth]{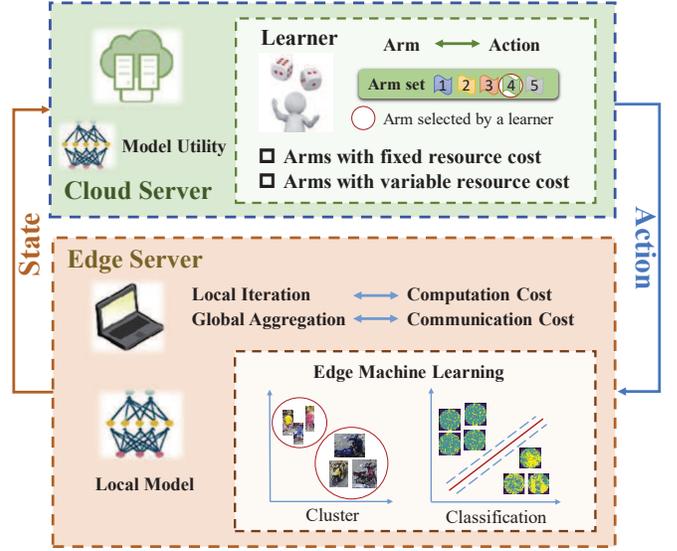}
  }
  \center\caption{Multi-armed Bandit Algorithm based \textsf{OL4EL}. \label{fig: ol for fl}}
\end{figure}

With our budget-limited bandit formulation, we propose an Online Learning for Edge Learning (\textsf{OL4EL}) algorithmic framework, which is shown in Figure \ref{fig: ol for fl}, for the Cloud server to determine the optimal sequence of \textit{global update interval (i.e., arm)} for either synchronous or asynchronous edge learning under given resource constraints (i.e., budgets), where the resource consumption (i.e., arm cost) could be either fixed or variable.

\textbf{1) Fixed resource cost:}
We assume that both the computation resource cost per local iteration and the communication resource cost per global update are fixed as constants during the entire learning process.
Then the resource cost of each candidate \textit{arm (global update interval)} can be directly calculated.
In this case, inspired by a budget-limited MAB solution in~\cite{tran2012knapsack}, we design an OL strategy for the Cloud server to determine the approximate optimal arm on-the-fly for edge servers by calculating the arm that provides the highest upper confidence bound of the estimated expectation of the learning utility while satisfying resource constraints.

The \textsf{OL4EL} mainly consists of two phases: initialization and dynamic decision.
In the initial phase, the Cloud server tries each feasible arm to coordinate the edge servers and measures the learning utility under each arm.
After initialization, the Cloud server enters into the dynamic decision phase and performs the following steps for each edge server at each slot:
\begin{itemize}
\item \textit{Utility-cost ordering:} Sort the candidate arm set in a descent order according to the current estimated utility per cost, which is the ratio between the expected learning utility and resource cost of each arm.
\item \textit{Frequency calculation:} Calculate the maximal frequency of each candidate arm supposing it is the only feasible arm, without exceeding the residual resource cost.
\item \textit{Probabilistic selection:} Randomly choose an arm in the candidate set with a probability proportional to the frequency of each arm.
\end{itemize}

The chosen arm is then the current approximate optimal global update interval for the corresponding edge server.

\textbf{2) Variable resource cost:}
In practice, the consumption rates for both types of resource evolve with the concurrent workloads or dynamic environments of edge servers.
In this case, the cost of different arms can be considered as i.i.d. random variables with different expectations \cite{ding2013multi}.
The similar idea of upper confidence bound algorithm in the case of fixed cost can be adopted.
However, the Cloud server needs to not only explore the learning utility of an arm, but also its resource cost.
Therefore, similar algorithm procedures in the fixed case can be migrated here, except for that, in the aforementioned \textit{Utility-cost ordering} step, the utility per cost of each arm is calculated as the ratio between the expected learning utility and the expected resource cost.

Considering the discussions above, \textsf{OL4EL} algorithms for both synchronous and asynchronous edge learning can be easily achieved.
The main difference is that the Cloud has to maintain only one bandit model for all edge servers in synchronous EL but different bandit models for all edge servers in asynchronous EL.

\section{Performance Evaluation}\label{sec:Experi}

\subsection{Setup}

\textbf{Testbed experiments and simulations.}
We encapsulated the Java codes of \textsf{OL4EL} on both edge servers and the Cloud using docker containers, which were deployed in an edge-cloud testbed composed of three mini PCs as edge servers and a workstation as the Cloud server.
Besides, to further investigate the performance of \textsf{OL4EL} in large scale systems, we constructed a simulator composed of a cloud server and 3 to 100 edge servers with the same \textsf{OL4EL} deployment.
We compared \textsf{OL4EL} with two baseline methods including the distributed training with fixed update intervals $I$ (referred to as \textsf{Fixed I}) and the state-of-the-art synchronous EL algorithm in~\cite{wang2018edge} (referred to as \textsf{AC-sync}).
\textsf{OL4EL-sync} and \textsf{OL4EL-async} denote synchronous and asynchronous \textsf{OL4EL} methods, respectively.

\textbf{Learning models and datasets.}
To demonstrate the compatibility of our proposed \textsf{OL4EL},
We adopted K-means based clustering and Support Vector Machine (SVM) based classification as unsupervised and supervised learning tasks for our edge learning.
Evaluations with K-means are set to cluster a real-world \textsf{traffic image} dataset containing $20,000$ images clipped from surveillance videos on YouTube Live into $K=3$ clusters.
Evaluations with SVM based classification are conducted on a real-world \textsf{wafer image} dataset, in which we considered 59-dimensional features with $20,000$ wafer images in smart manufacturing and tags of 8 classes.

\textbf{Evaluation metrics.}
To measure the learning performance, we collected the F1 score and prediction accuracy as the evaluation metrics for K-means and SVM, respectively.
We use time as the resource metric and define the \textit{resource constraint} as a given time budget (e.g., remaining time of battery or service) for each edge server. Then, the computation and communication resource cost refer to the execution time for local iterations, and the edge-Cloud communication duration for the updates, respectively.
In testbed experiments, they are measured as the practical system time cost during iterations or updates.
In simulations, they are assigned with different integers representing corresponding units of time for each iteration or update, and calculated according to the number of executions.
Time is measured in units of milliseconds (ms) in the experiments.

\begin{figure}
  \centering
  \subfigure{
        \includegraphics[width=0.5\textwidth]{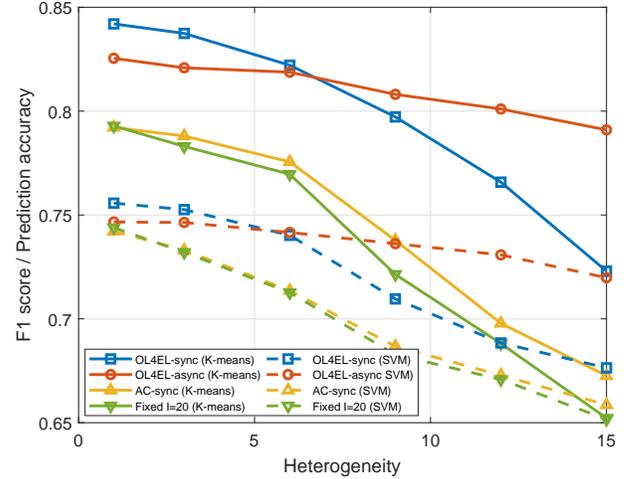}
  }
  \center\caption{Model Accuracy vs. Heterogeneity. \label{fig: The result of k-means and SVM2}}
\end{figure}

\subsection{Results}

\textbf{1) Impact of edge heterogeneity on model accuracy:}
Figure~\ref{fig: The result of k-means and SVM2} compares \textsf{OL4EL} with two comparison algorithms on heterogeneous edge servers using testbed experiments with fixed resources $5,000$ ms.
The heterogeneity of edge servers is measured as the ratio of processing speed of the fastest edge server to that of the slowest one.
Particularly, $H=1$ indicates the case of fully homogeneity among edge servers.

We can see that, the model accuracy of all algorithms falls when the heterogeneity increases, since larger heterogeneity will lead to less aggregations on the Cloud.
Nonetheless, our algorithms \textsf{OL4EL} significantly outperform other two comparison algorithms \textsf{AC-sync} and \textsf{Fixed I}.
When the heterogeneity is lower (i.e., $H \leq 5$), \textsf{OL4EL-sync} demonstrates higher accuracy in both F1 score and prediction than that of \textsf{OL4EL-async}, which is because the synchronous strategy has no stale updates from less heterogeneous edges.
However, when the heterogeneity is higher, \textsf{OL4EL-async} then shows great superior and the reason is that asynchronous architecture ensures a higher efficiency that fast edge servers can immediately update the global model without waiting for the others.
Specifically, our asynchronous algorithm \textsf{OL4EL-async} can achieve at most 12$\%$ higher accuracy than \textsf{AC-sync} and \textsf{Fixed I}.
It is worth to be noted that our synchronous algorithm \textsf{OL4EL-sync} also outperforms \textsf{AC-sync} even the heterogeneity is high as all computations of \textsf{OL4EL-sync} are performed on the Cloud, resulting in less edge resource consumption than \textsf{AC-sync} that requires local calculations at edge servers.

\begin{figure}
  \centering
  \subfigure{
        \includegraphics[width=0.5\textwidth]{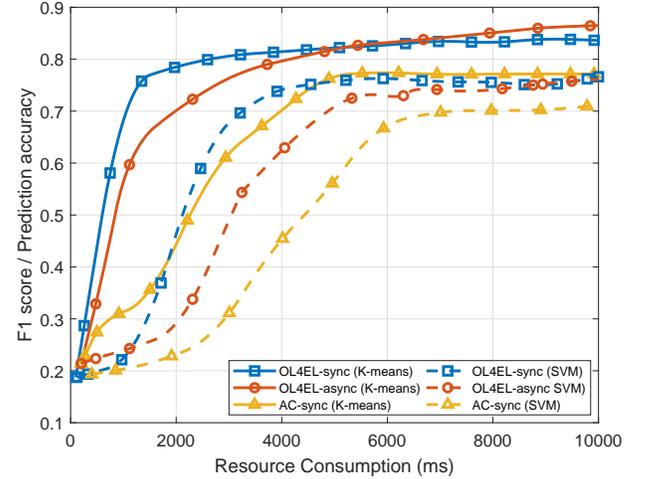}
  }
  \center\caption{Model Accuracy vs. Edge Resource Consumption. \label{fig: The result of k-means and SVM22}}
\end{figure}

\textbf{2) Trade-off between model accuracy and resource consumption:}
Figure~\ref{fig: The result of k-means and SVM22} shows the testbed experimental results on the long-term performance of \textsf{OL4EL} versus the edge resource consumption, under the edge heterogeneity as $6$.

In Figure~\ref{fig: The result of k-means and SVM22}, with the increase of resource consumption, all algorithms gradually achieve better model accuracy, which demonstrates the intrinsic trade-off between learning performance and resource consumption.
Under any resource consumption, \textsf{OL4EL-sync} and \textsf{OL4EL-async} can always achieve higher accuracy than the baseline method \textsf{AC-Sync}, which depicts the better trade-off between learning performance and resource consumption of \textsf{OL4EL}.
Particularly, when more resource is consumed, \textsf{OL4EL-async} will get the highest model accuracy. The reason is that asynchronous architecture allows more global updates, thus significantly improves the learning efficiency.

\begin{figure*}
  \centering
  \subfigure[K-Means]{
        \includegraphics[width=8cm]{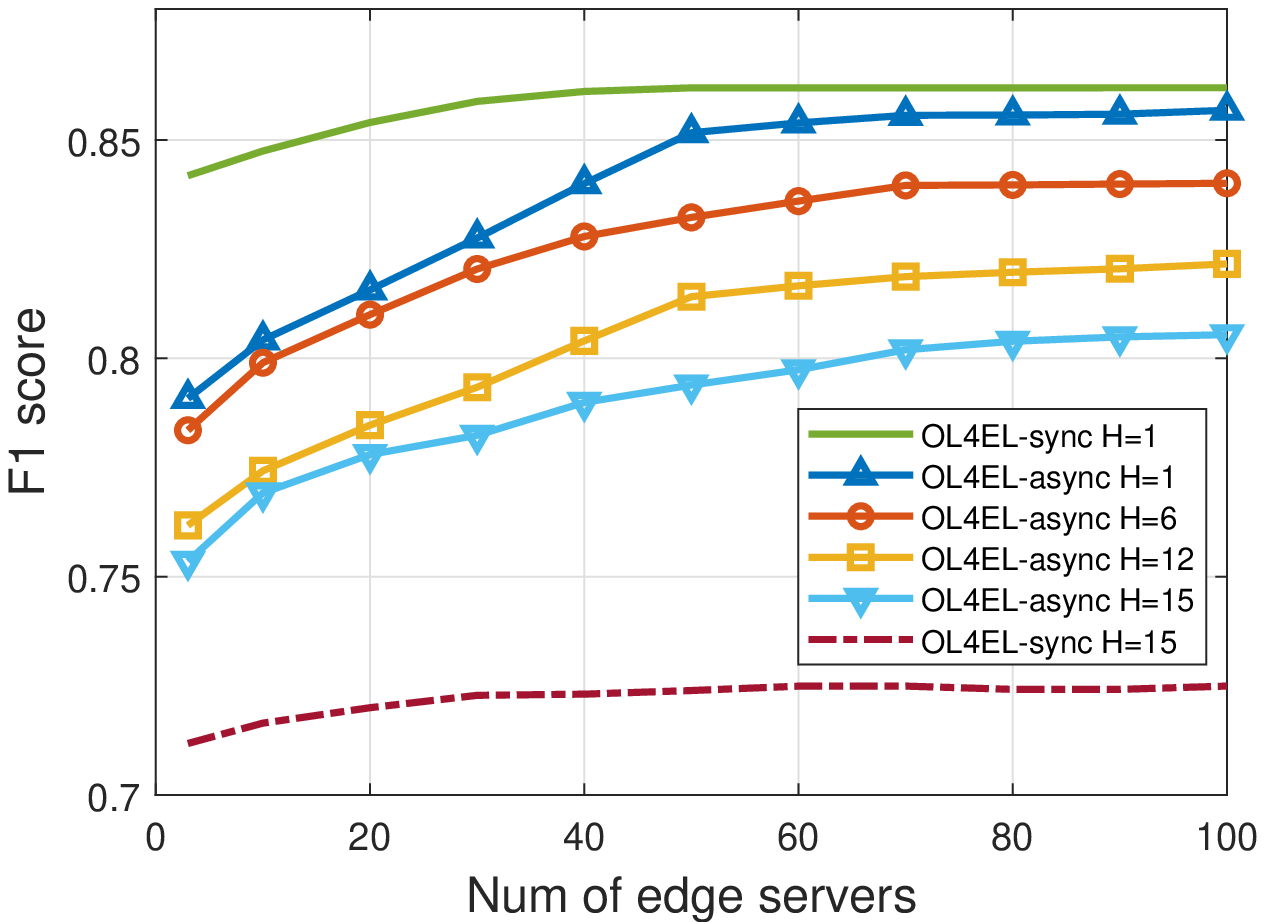}
  }
  \subfigure[SVM]{
        \includegraphics[width=8cm]{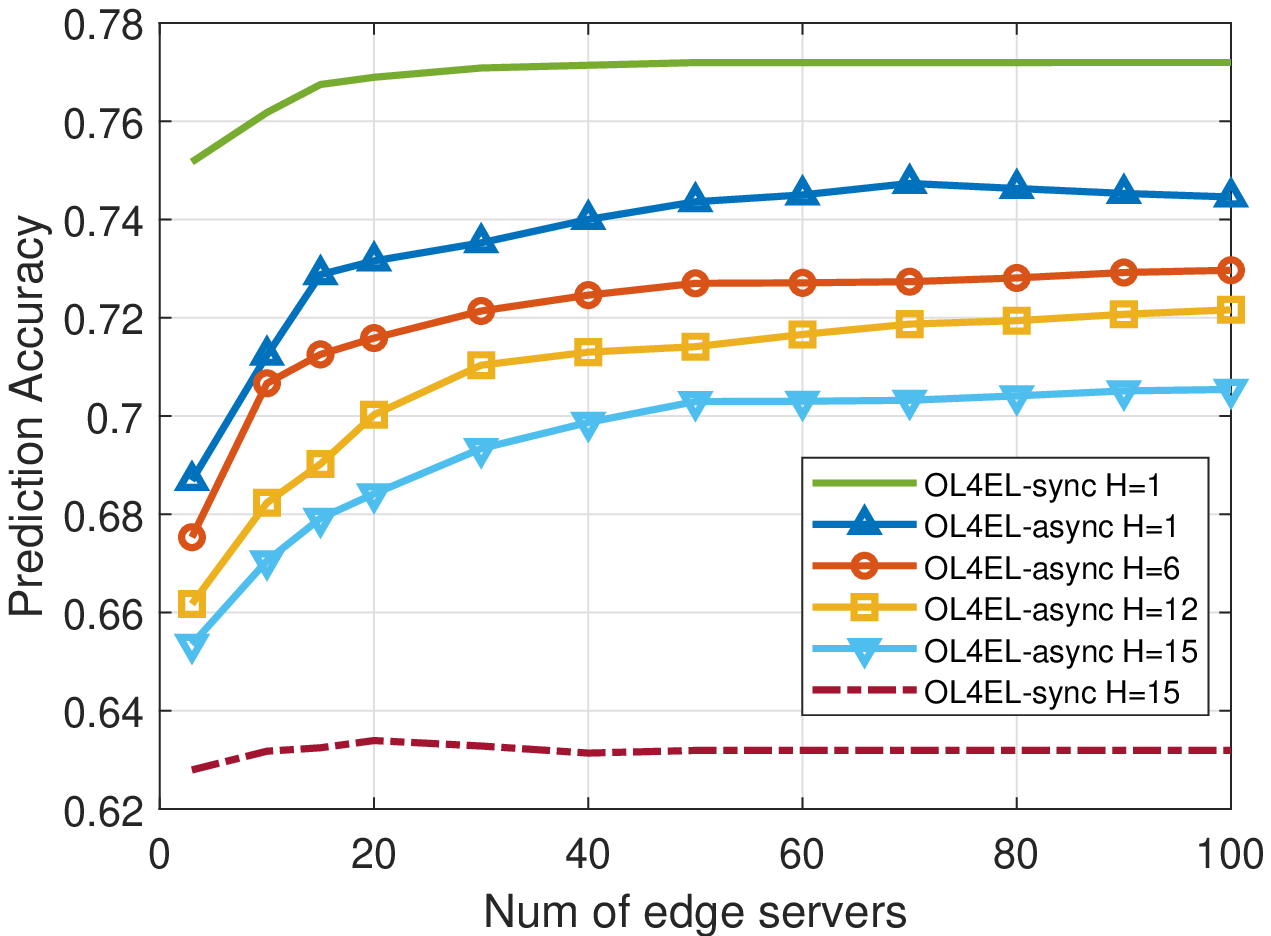}
  }
  \caption{Model Accuracy vs. Number of Edge Servers: a) the result of K-means; b) the result of the SVM. \label{fig: The result of kmeans}}
\end{figure*}

\textbf{3) Impact of edge number on model accuracy:}
To study the scalability of \textsf{OL4EL} algorithm, we conducted simulations with different numbers of edge servers ranging from $3$ to $100$.
Figure~\ref{fig: The result of kmeans} demonstrates the model accuracy of \textsf{OL4EL-async} with the increase of edge numbers under variable edge heterogeneity.

As shown, as the number of edge servers increases, the performance of \textsf{OL4EL-async} becomes better, since that more information is aggregated for model updates.
In addition, with the increase of heterogeneity of edge servers, the model accuracy of both K-means and SVM become worse, which is consistent with results in Figure~\ref{fig: The result of k-means and SVM2}.
The reason is that updates from slower edge servers decreases with the increase of heterogeneity, thus contributing less to global model training.
Moreover, we also compare \textsf{OL4EL-async} with \textsf{OL4EL-sync} in Figure~\ref{fig: The result of kmeans}.
When edge servers are homogeneous (i.e. $H=1$), \textsf{OL4EL-sync} achieves the best performance since all edge servers can be utilized to update the global model. However, as the heterogeneity increase, \textsf{OL4EL-sync} degrades dramatically.
For example, when $H=15$, it performs even worse than \textsf{OL4EL-async} because the model updates are determined by the slowest edge.

\section{Conclusion}\label{sec:Conclu}

In this article, we investigate how to use online learning to optimize the decision making for efficient edge-cloud collaborative learning among heterogeneous edges under limited resources.
We develop an algorithm named \textsf{OL4EL} that can support both synchronous and asynchronous learning patterns.
By using real-world datasets, and K-means and SVM as unsupervised and supervised learning tasks respectively, we conducted extensive simulations and testbed experiments based on docker containers to evaluate the performance of \textsf{OL4EL}.
Evaluation results demonstrated that \textsf{OL4EL} achieves 12$\%$ higher model accuracy than the state-of-the-art algorithms.

%

\small
\bibliographystyle{IEEEtran}

\section*{Biography}
\begin{IEEEbiographynophoto}
{Qing Han} (qinghan@stu.xjtu.edu.cn) received her M.Sc. degree from Xi'an Jiaotong University (XJTU) of China in 2018. She is currently working towards the Ph.D. degree in the School of Computer Science and Technology at XJTU. Her research interests include edge computing and federated learning.
\end{IEEEbiographynophoto}

\begin{IEEEbiographynophoto}
{Shusen Yang} (shusenyang@mail.xjtu.edu.cn) received the Ph.D. degree from Imperial College London in 2014. He is a professor and director of the National Engineering Laboratory for Big Data Analytics, and deputy director of Ministry of Education(MoE) Key Lab for Intelligent Networks and Network Security, both at Xi'an Jiaotong University. 
His research interests include mobile networks, networks with human in the loop, data-driven networked systems and edge computing.
\end{IEEEbiographynophoto}

\begin{IEEEbiographynophoto}
{Xuebin Ren} (xuebinren@mail.xjtu.edu.cn) received his Ph.D. degree from Xi'an Jiaotong University (XJTU) of China in 2017. He has been a visiting Ph.D. student at Imperial College in 2016. He is currently an assistant professor in School of Computer Science and Technology, and the National Engineering Laboratory for Big Data Analytics, both at XJTU. His research interests focus on data privacy protection, federated learning and privacy-preserving machine learning.
\end{IEEEbiographynophoto}

\begin{IEEEbiographynophoto}
{Cong Zhao} (c.zhao@imperial.ac.uk) received his Ph.D. degree from Xi'an Jiaotong University (XJTU) in 2017. He is currently a Research Associate in the Department of Computing at Imperial College London. His research interests include edge computing, meta learning, computing economics, and people-centric sensing.
\end{IEEEbiographynophoto}

\begin{IEEEbiographynophoto}
{Jingqi Zhang} (zhangjingqi@stu.xjtu.edu.cn) received her B.Sc. degree from Xi'an Jiaotong University of China in 2018. She is currently a master candidate at Xi'an Jiaotong University. Her research interests include big data analysis and edge learning.
\end{IEEEbiographynophoto}

\begin{IEEEbiographynophoto}
{Xinyu Yang} (yxyphd@mail.xjtu.edu.cn) received his B.Sc., M.Sc., and Ph.D. degrees in computer science and technology from Xi'an Jiaotong University (XJTU), China, in 1995, 1997, and 2001, respectively. He is currently a Professor in the School of Computer Science and Technology, XJTU. His research work focuses on distributed systems and artificial intelligence.
\end{IEEEbiographynophoto}

\end{document}